\documentclass[final,3p,times,sort&compress]{elsarticle}

\usepackage{multirow,setspace,times,amssymb,amsmath,graphicx,color,rotating,subfigure,url}
\usepackage{lineno}
\usepackage{natbib}
\usepackage{subfigure}
\usepackage{lscape}
\usepackage{multirow}
\usepackage{booktabs}
\usepackage{threeparttable}   
\begin{document}

\begin{frontmatter}

\title{Testing the performance of technical trading rules in the Chinese market}
\author[BS]{Shan Wang}
\author[BS,RCE]{Zhi-Qiang Jiang\corref{cor}}
\ead{zqjiang@ecust.edu.cn}
\author[RCE,AS]{Sai-Ping Li}
\author[BS,RCE,SS]{Wei-Xing Zhou\corref{cor}}
\ead{wxzhou@ecust.edu.cn} %
\cortext[cor]{Corresponding authors.}

\address[BS]{Department of Finance, School of Business, East China University of Science and Technology, Shanghai 200237, China}
\address[RCE]{Research Center for Econophysics, East China University of Science and Technology, Shanghai 200237, China}
\address[AS]{Institute of Physics, Academia Sinica, Nankang, Taipei 11529, Taiwan}
\address[SS]{Department of Mathematics, School of Science, East China University of Science and Technology, Shanghai 200237, China}

\begin{abstract}
  Technical trading rules have a long history of being used by practitioners in financial markets. Their profitable ability and efficiency of technical trading rules are yet controversial. In this paper, we test the performance of more than seven thousands traditional technical trading rules on the Shanghai Securities Composite Index (SSCI) from May 21, 1992 through June 30, 2013 and Shanghai Shenzhen 300 Index (SHSZ 300) from April 8, 2005 through June 30, 2013 to check whether an effective trading strategy could be found by using the performance measurements based on the return and Sharpe ratio. To correct for the influence of the data-snooping effect, we adopt the Superior Predictive Ability test to evaluate if there exists a trading rule that can significantly outperform the benchmark. The result shows that for SSCI, technical trading rules offer significant profitability, while for SHSZ 300, this ability is lost. We further partition the SSCI into two sub-series and find that the efficiency of technical trading in sub-series, which have exactly the same spanning period as that of SHSZ 300, is severely weakened. By testing the trading rules on both indexes with a five-year moving window, we find that the financial bubble from 2005 to 2007 greatly improve the effectiveness of technical trading rules. This is consistent with the predictive ability of technical trading rules which appears when the market is less efficient.
\end{abstract}

\begin{keyword}
   Econophysics \sep Technical analysis \sep Data-snooping \sep Bootstrap method  \sep Superior predictive ability
\end{keyword}

\end{frontmatter}

%

\section{Introduction}

Technical trading rules have been widely used to detect the market trends for financial practitioners. In academia, numerous studies have been conducted to try to answer the question whether technical strategies are useful through applying trading rules to different financial markets, but instead give conflicting conclusions. On the one hand, some scholars advocate that technical rules do have predictive ability to earn excess profits. Treynor and Ferguson found that it is possible to get the abnormal profits with the past prices and other valuable information \cite{Treynor-Ferguson-1985-JF}. Brock et al showed that in the U.S. market, technical trading rules could reveal some certain return patterns when applied to the Dow Jones Industrial Average (DJIA) \cite{Brock-Lalonishok-Lebaron-1992-JF}. In the European monetary system, Neely et al reported that technical trading rules could be useful even with the use of out-of-sample test \cite{Neely-Weller-1999-JIMF}. On the other hand, technical rules are found to be useless by other researchers. Lucke carried out a test on the effectiveness of head-and-shoulder trading rules on foreign exchange markets and found that returns to head-and-shoulder trading rules were not significantly positive \cite{Lucke-2003-AE}. Anderson and Faff found that in futures markets, the profit of technical trading rules was not significantly obtainable \cite{Anderson-Faff-2005-ARJ}. To investigate the efficiency of technical trading rules, Kung focused on profits produced by buying signals and selling signals, using the Taiwan Stock Exchange Weighted Index and found that for these rules, returns from buy signals are higher than those from sell signals. The predictive power became less effective over 1997-2007 when compared with periods 1975-1985 and 1986-1996, indicating an improvement of efficiency of the Taiwan stock market. \cite{Kung-2009-RAE}. Fang used the out-of-sample test, and found that no technical trading rule had the predictive ability \cite{Fang-Jacobsen-Qin-2014-RFE}. Obviously, the profitability of technical trading strategies depends on the states of the market and the choice of specific trading rules. Yu et al argued that the predictive ability of technical trading rules appears only when the market is less efficient \cite{Yu-Nartea-Gan-Yao-2013-IREF}. Hudson et al tested the possibility of earning excess returns by using technical analysis in the UK market and found that although the technical trading rules do have predictive ability, it considerably weakens when one takes into account trading costs \cite{Hudson-Dempsey-Keasey-1996-JBF}.

The effectiveness of technical trading rules is further in doubt due to the data-snooping effect, which occurs when the same data is used more than once for the purpose of inference or model selection. If the best model is obtained by an extensive specification search, it will be highly possible that this good model is chosen by luck rather than its actual forecasting ability \cite{White-2000-Em}. The existence of dependence between all the tested trading rules imply that the data-snooping effect would inevitably amplify the significance levels of conventional hypothesis tests. Jensen and Bennington demonstrated that data-snooping on the performance of technical trading rules was a selection bias \cite{Jensen-Bennington-1970-JF}. Lo and Mac Kinlay showed that effects of data-snooping were substantial by using analytical calculation and simulations \cite{Lo-MacKinlay-1990-RFS}. Brock et al also recognized the existence of data-snooping and evaluate by fitting several models to the raw data and created new sample series by resampling the residuals \cite{Brock-Lalonishok-Lebaron-1992-JF}. They applied 26 simple trading rules, including moving average and range break-out rules, to the US stock markets and found that technical analysis was indeed capable of providing significant economic contents.

In order to correct the biased effects induced by data-snooping, a statistical test, called ``White's Reality Check'' (abbreviated as WRC), was proposed to examine whether the profit of technical trading rules is true or just from luck \cite{White-2000-Em}. Sullivan el al also applied this test to check whether the technical strategies ($7846$ trading rules) had the predictive power in DJIA and S\&P 500 index and found that the superior performance of the best trading rule could beat the benchmark, but could not repeat in the out-of-sample experiment \cite{Sullivan-Timmermann-White-1999-JF}. Chen et al tested the same trading strategies as in Ref.~\cite{Sullivan-Timmermann-White-1999-JF} in Asian stock markets and found that the WRC $p$-values of different markets were not the same. They also found that the predictive ability in Asian stock markets was not as good as that in the US.

Due to the deficiency of WRC tests whose $p$-value would increase when bad trading rules produced negative performance measurements, the Superior Predictive Ability (SPA) test was proposed to check the existence of predictive ability of a superior model \cite{Hansen-2005-JBES}. Hsu and Kuan reexamined the profitability of technical trading rule set in four indexes by using both WRC and SPA tests and found that the trading rules could only earn profits in young markets (NASDAQ Composite and Russell) rather than in developed markets (DJIA and S\&P 500). Hsu et al further extended the WRC and SPA test into a stepwise SPA test and a stepwise Reality Check test. With these two extended tests, they examined the predictive ability of technical trading rules in emerging markets, and demonstrated that technical trading rules do have significant predictive abilities. Park et al used both WRC and SPA tests to investigate the profitability of technical trading rules in U.S futures markets and found that the best trading rules generated statistically significant economic profits only for some futures contracts \cite{Park-Irwin-2010-JFutM}. This suggests that technical trading rules do not generally have the profitable ability in U.S. futures markets when take into account data-snooping biases. Shynkevich applied the technical trading rules to the growth and small cap segments of the US equity market, and found that mechanical trading strategies had lost their predictive ability when data-snooping was considered \cite{Shynkevich-2012-JBF}.

In this paper, we will apply the 7846 trading rules into Shanghai Securities Composite Index (SSCI) and Shanghai Shenzhen 300 Index (SHSZ 300) in the Chinese stock markets to study whether any of these technical trading rules is capable of making profits under the consideration of the data-snooping effect. This paper is arranged as follows. In Section 2, we introduce the data information. All the trading rules and the benchmark are presented in Section 3. In Section 4, we will describe details of the SPA test. In Section 5, we will present the empirical results. Section 6 is the conclusion.

\section{Data Sets}

We apply the trading rules in our strategy pools on two important indexes (SSCI and SHSZ 300) in Chinese stock markets to investigate which technical strategy has the best performance by means of the SPA tests. Both daily indexes are provided by the financial data company RESSET and China Investment Security database. The SSCI covers a period from May 21, 1992 through June 30, 2013, which leads to 5144 data points. SHSZ 300 spans a period ranging from April 8, 2005 and June 30, 2013, which results in 1996 data points. We estimate the daily logarithmic difference return and find that the largest daily return is $28.86\%$ for SSCI and $8.93\%$ for SHSZ 300. The smallest return for SSCI and SHSZ 300 is  $-17.91\%$ and $-9.70\%$. We also find that there are 32 returns whose values are larger than $10\%$ or less than $-10\%$ for SSCI. All these returns occurred before Dec 16, 1996 when the price limit rule was launched in Chinese markets. We also estimate the skewness and kurtosis of daily returns for both indexes. The skewness of the SSCI returns is 1.18, which indicates that the return distribution is right skewed. For SHSZ 300, the return distribution is a little bit left skewed and the corresponding skewness is -0.37. The kurtosis values of the returns for both indexes are greater than 3, which is a typical value of normal distributions. For the SSCI, the kurtosis is 22.07, which is about four times greater than the kurtosis 5.71 for SHSZ 300. This means that the return distribution possesses the characteristic of leptokurtic.

\section{Technical Trading Rules and Benchmark}

To begin with, it is necessary to specify all the technical trading rules which will be tested. Brock et al have used 26 simple trading rules to test the performance of technical trading strategies in the U.S. market \cite{Brock-Lalonishok-Lebaron-1992-JF}. Sullivan et al. extended these simple trading rules to a larger universal technical analysis space \cite{Sullivan-Timmermann-White-1999-JF}, which were also tested in Ref.~\cite{Chen-Huang-Lai-2009-JAsianE}. Our analysis is also based on this universal technical analysis space, which comprises 7846 trading rules from five main technical analysis catalogs, namely filter rules, moving averages, support and resistance, channel breakouts, and on-balance volume averages.

\subsection{Filter Rules}

The standard filter rule was explained in Ref.~\cite{Fama-Blume-1965-JB}. The $x$ percent filter is defined as follows. The stock should be bought when the daily closing price moves up by at least $x$ percent, and the position will be held until its price moves down at least $x$ per cent from the subsequent high, and simultaneously the investor should sell the stock and go to a short position. This short position is maintained until the daily closing price rises by at least $x$ percent above the subsequent low, at which time the investor covers the short position and buys the security. Any movement less than $x$ percent in either direction is ignored.

We rely on four parameters to implement the filter rules: (1) a change of price to initiate short or long positions $x$ ($0.005$, $0.01$, $0.015$, $0.02$, $0.025$, $0.03$, $0.035$, $0.04$, $0.045$, $0.05$, $0.06$, $0.07$, $0.08$, $0.09$, $0.1$, $0.12$, $0.14$, $0.16$, $0.18$, $0.2$, $0.25$, $0.3$, $0.4$, $0.5$, ranging from 0.005 to 0.5, giving a total of $24$ values) ; (2) a change of price to clear a position $b$ ($0.005$, $0.01$, $0.015$, $0.02$, $0.025$, $0.03$, $0.04$, $0.05$, $0.075$, $0.1$, $0.15$, $0.2$ giving a total of $12$ values), which means that when the price goes down (up) by $b$ percent, the long or short positions will be cleared; (3) a subsequent extremum, which is the most recent closing price that is less (greater) than the $e$ previous closing prices ($1$, $2$, $3$, $4$, $5$, $10$, $15$, $20$, giving a total of $8$ values); (4) the number of days $c$ a position is held ($5$, $10$, $25$ or $50$ days, giving a total of $4$ values). All these parameters together give a total of $497$ filter rules.

\subsection{Moving Averages}

Moving average is another very popular technical strategy. A buy (sell) signal is generated when the short-term average price crosses the long-term average price from below (above). We need five parameters to implement the procedure of moving average rules: (1) a number of days used to calculate the short-term and long-term moving average $n$  (ranging from $2$ to $250$, namely $2$, $5$, $10$, $15$, $20$, $25$, $30$, $40$, $50$, $75$, $100$, $125$, $150$, $200$, $250$, giving a total of $15$ values); (2) a predefined band $b$ to ensure that the short-term moving average is greater or less enough than the long-term moving average to generate signals ($0.001$, $0.005$, $0.01$, $0.015$, $0.02$, $0.03$, $0.04$, $0.05$, giving a total of $8$ values), which can reduce the noisy signals when the stock is in turbulent state; (3) a number of days for the signals remain valid before action is taken ($2$, $3$, $4$ or $5$ days, giving a total of $4$ values); (4) a number of days that a position should be held $c$ ($5$, $10$, $25$ or $50$ days, giving a total of $4$ values). In addition, $9$ MA rules in Ref.~\cite{Brock-Lalonishok-Lebaron-1992-JF}, which combines the short-term MA of $1$, $2$, and $5$ days and long-term MA of $50$, $150$, $200$ days, together with $1$ percent band and $10$ days holding period, are added. All of these result in $2049$ trading rules.

\subsection{Support and Resistance }

The support and resistance strategy is to buy (sell) when the closing price upward (downward) exceeds the maximum (minimum) closing price over the previous $n$ days ($5$, $10$, $15$, $20$, $25$, $50$, $100$, $150$, $200$, $250$, giving a total of $10$ values). An alternative way is to use the most recent closing price that is greater or less than the $e$ previous closing price as maximum or minimum ($2$, $3$, $4$, $5$, $10$, $20$, $25$, $50$, $100$, $200$, giving a total of $10$ values). Like the moving average, a fixed percentage filter band $b$ ($0.001$, $0.005$, $0.01$, $0.015$, $0.02$, $0.03$, $0.04$, $0.05$, giving a total of $8$ values), a time delay filter $d$ days ($2$, $3$, $4$ or $5$ days, giving a total of $4$ values) and a fixed holding time $c$ days ($5$, $10$, $25$ or $50$ days, giving a total of $4$ values) can be included. These result in $1220$ rules.

\subsection{Channel Breakouts}

A stock movement channel can be defined when the difference of high and low over the previous $n$ days ($5$, $10$, $15$, $20$, $25$, $50$, $100$, $150$, $200$, $250$, giving a total of $10$ values) is within $x$ percent ($0.005$, $0.01$, $0.02$, $0.02$, $0.05$, $0.075$, $0.1$, $0.15$, giving a total of $8$ values) of the low. A buy (sell) signal is then generated when the closing price moves up (down) the upper (lower) channel. When a filter band is included, which means the change should not be less than $b$ percent ($0.001$, $0.005$, $0.01$, $0.015$, $0.02$, $0.03$, $0.04$, $0.05$, giving a total of $8$ values), and a holding position $c$ days ($5$, $10$, $25$ or $50$ days, giving a total of $4$ values), more rules would be generated. All of these permutations will produce $2040$ rules in total.

\subsection{On-balance Volume Averages}
The on-balance volume averages strategy is based on the moving average rules. Here, the on-balance volume indicator ($OBV$) rather than the stock price is applied to the moving average method. A buy or sell signal is then generated the same way as the moving average rules. $OBV$ is calculated by keeping a running total of the indicator each day and adding (subtracting) the entire amount of daily volume when the closing price increases (decreases). We have a total of $2040$ rules in this category.

\subsection{Benchmark}

A benchmark is set to find out a trading rule outperforming the market. Different empirical studies use different criteria to define the benchmark. Following what is used in Ref.~\cite{Brock-Lalonishok-Lebaron-1992-JF}, our benchmark is out of market.

\section{Superior Predictive Ability Test}

In most of the existing work in the literature on technical trading rules, the most profitable trading rules are usually obtained by applying a pool of trading rules to past data. As a result, some of the trading rules may be selected as useful just by chance rather than their actual forecasting ability, which is mostly a result due to the data-snooping effect. The most popular testing methods taking into account to correct the data-snooping effect are WRC tests \citep{White-2000-Em} and SPA tests \citep{Hansen-2005-JBES}. Note that SPA tests are modified from WRC tests to offer more reliable calculations. In order to test whether the technical rules have the capacity to predict and produce profits after accounting for the effects of data-snooping in Chinese stock markets, we here choose the SPA test to perform our statistical tests.

\subsection{Definition of performance measurements}

There are two usual measurements of the performance of trading rules, namely the return and the Sharpe ratio.  For the return, the performance statistic of $k$-th trading rule can be defined as,
\begin{equation}
\langle f^r_k \rangle = \frac{1}{n}\sum^{T}_{t=R} f^r_{k,t+1},
\end{equation}
where $k$ means the $k$-th trading rule and the total number of technical trading rules is $l$. $n$ is the number of trading days from $R$ to $T$. $R$ equals to 250, since some trading rules require 250 data points before the trading day under consideration to implement, for example, the moving average rules.  We thus have $n=T-R+1$. $f_{k,t+1}$ represents the performance of the $k$-th trading rule when compared with the benchmark on day $t+1$, which is defined as
\begin{equation}
f^r_{k,t+1} = \ln (1+ r_{t+1} I_{k,t+1}) - \ln(1+ r_{t+1} I_{0,t+1} ),
\end{equation}
where $r_{t+1}$ is the relative return on day $t+1$, defined as $r_{t+1} = (p_{t+1} - p_t)/p_t$. $I_{k, t+1}$ and $I_{0, t+1}$ are the indication of market positions on day $t+1$ which are translated from the trading signals triggered from the trading rule $k$ and the benchmark respectively. If short selling is allowed, the possible value of the dummy variable $I$ could be $1$, $0$, and $-1$, representing the long, neutral, and short positions in the markets. If short selling is not allowed, the dummy variable $I$ could only take two values $1$ and $0$, representing the long and other positions in the market. The benchmark $I_0$ could be chosen as 1 and 0 to stand for the buy-and-hold strategy and out of the market. In the present paper, we only account for the situation with short selling.
We can also define the performance of trading rules based on the Sharpe ratio,
\begin{equation}
\langle f^s_k \rangle = \frac{1}{n}\sum^{T}_{t=R} f^s_{k,t+1},
\end{equation}
in which case $f^s_{k,t+1}$ could be defined as follows,
\begin{equation}
f^s_{k,t+1} = \frac{r_{k, t+1} - r_{f, t+1}}{\sigma_k} - \frac{r_{0, t+1} - r_{f, t+1}}{\sigma_0},
\end{equation}
where $\sigma_k$ is the variance of the daily return series $r_{k, t}$ generated by the $k$-th trading rule, $\sigma_0$ is the variance of the benchmark returns, and $r_{f, t+1}$ is the risk-free interest rate on day $t+1$. The return series $r_{k,t+1}$ of trading rule $k$ is formulated as follows,
\begin{equation}
 r_{k, t+1} = r_{t+1} I_{k, t+1},
\end{equation}
where $r_{t+1}$ is the daily return of the index series on day $t+1$ and $I_{k,t}$ is a dummy variable, indicating the market positions converted from the trading signals of the $k$-th trading rule. In our statistical tests, we set the benchmark to be zero for both performance measurements.

\subsection{Null hypothesis}

To test whether there is a technical trading rule that outperforms the benchmark, one can give the null hypothesis that the performance of the best trading rules is worse than that of the benchmark strategy. Based on the performance measurements, we can formulate the null hypothesis of SPA tests for the return $H^r_0$ and the Sharpe ratio $H^s_0$ as follows,
\begin{equation}
H_{0}^r : \max_{k=1,2,...l} \{ \langle f^s_{k} \rangle \}\leq 0 ~~~~{\rm{and}} ~~~~ H_{0}^s : \max_{k=1,2,...l} \{ \langle f^s_{k} \rangle \} \leq 0.
\end{equation}
Rejection of the null hypothesis implies that the best one from the whole technical trading rules can outperform the benchmark.

\subsection{Bootstrap method}

In order to test the null hypothesis, we adopt the stationary bootstrap method  \cite{Politis-Romano-1994-JASA} to generate the synthetic data $\langle f^r_k \rangle^* $ and $\langle f^s_k \rangle^*$  to estimate the $p$-value for both performance measurements. The synthetic data $\langle f^r_k \rangle^*$ and $\langle f^s_k \rangle^*$ can be determined as follows,
\begin{equation}
\langle f_k \rangle^* =\frac{1}{n}\sum^{T}_{t=R} f^{*}_{k,t+1},
\end{equation}
where $f^{*}_{k, t}$ is obtained by block shuffling the series $f_{k, t}$. Notice that the superscripts $r$ and $s$ of $f$ are omitted here.

For the block shuffling process, the size of shuffled blocks is determined by the ``smoothing parameter'' $q$ in the range of $[0, 1]$, which gives the expected length of the blocks as $1/q$. For a given series $\theta(t)$ with $R \le t \le T$, the block resampling series $\theta^*(t)$ can be obtained through the following procedure,

\begin{enumerate}
\item Set $t=R$ and $\theta^*(t)=\theta(i)$, where $i$ is random, independently and uniformly drawn from $R,\ldots,T$.
\item Increasing $t$ by 1. If $t>T$, stop. Otherwise, draw $u$ from the standard uniform distribution $[0,1]$. If $u<q$, set $\theta^*(t) = \theta(i)$, where $i$ is random, independently and uniformly  drawn from $R,\ldots,T$. If $u\geq q$, set $\theta^*(t)=\theta(i+1)$; if $i+1 > T$, we reset $i=R$.
\item Repeat step two.
\end{enumerate}

\subsection{Estimating the $p$-value}

The statistic of SPA tests can be defined as follows \cite{Hansen-2005-JBES},
\begin{equation} \label{Eq:SPA:Statistics}
T_{l}=\max\left( \max_{k=1,...,l}\frac{n^{1/2}\langle f_{k} \rangle}{\widehat{\omega}_{k}},0 \right),
\end{equation}
where $\widehat{\omega}^{2}_{k}$ represents the estimator of the variance ${\rm{var}}(\sqrt{n} \langle f_{k} \rangle)$, which will be introduced in the following. Note that the designation $f$ in this subsection could be either $f^r$ or $f^s$.

In order to proceed on with the SPA test and to evaluate the $p$-value, we need to estimate the statistics for each bootstrap sample,
\begin{equation} \label{Eq:SPA:Statistics:Boostrap}
T_{l}^*=\max \left( \max_{k=1,...,l} \frac{ n^{1/2}\overline{Z}_{k} }{\widehat{\omega}_{k}},0 \right),
\end{equation}
where $\overline{Z}_{k}$ can be obtained through the following equation,
\begin{equation} \label{Eq:SPA:Statistics:Boostrap:Zbar}
\overline{Z}_{k} = 1/n \sum_{t=1}^{n}Z_{k,t},
\end{equation}
where $Z_{k,t}$ is written as,
\begin{equation} \label{Eq:SPA:Statistics:Boostrap:Z}
Z_{k,t}=f_{k,t}^*- g\left( \langle f_k \rangle \right),
\end{equation}
where $g(x)$ is defined as $g(x)=x \textit{\textbf{1}} \left( n^{1/2} x \geq - \widehat{\omega}_{k} \sqrt{2\ln\ln(n)} \right)$. Note that $\textit{\textbf{1}}(\cdot)$ is an indicator function which generates 1 if the expression in parentheses is true and 0 if the condition is not met.

As described above, one bootstrap sample will give one $T_{l}^*$. We can regenerate the block shuffling sample until there are 500 values of $T_{l}^*$. The bootstrap $p$ value could then be given as,
\begin{equation}
p=\sum_{i=1}^{500}\frac{\textit{\textbf{1}} (T^*_{l,i} > T_l )}{500}.
\end{equation}
The upper (respectively, lower) bounds of the $p$-value $p_{u}$ (respectively, $p_{l}$) can also be estimated by substituting the function $g(x)$ as $g(x)=x$ (respectively, $g(x) = \max(x, 0)$).

The $\widehat{\omega}_{k}$ in Eqs.~(\ref{Eq:SPA:Statistics}) and (\ref{Eq:SPA:Statistics:Boostrap}) can be obtained from the following procedure,
\begin{equation}
\widehat{\omega}^{2}_{k}=\widehat{\gamma}_{0,k}+2\sum_{t=1}^{n-1}\kappa(n,t)\widehat{\gamma}_{t,k},
\end{equation}
where
\begin{equation}
\widehat{\gamma}_{t,k}=n^{-1}\sum_{j=1}^{n-t}(f_{k,j}- \langle f_{k} \rangle )(f_{k,t+j}- \langle f_{k} \rangle), ~~~~t=0,1,...,n-1,
\end{equation}
are the empirical covariances. The kernel weights are given by the following form,
\begin{equation}
\kappa(n,t)=\frac{n-t}{n}(1-q)^{t}+\frac{t}{n}(1-q)^{n-t},
\end{equation}
where $q$ is the parameter in the block shuffling procedure.

\section{Empirical Test}

\begin{figure}[htb]
\centering
\includegraphics[width=16cm]{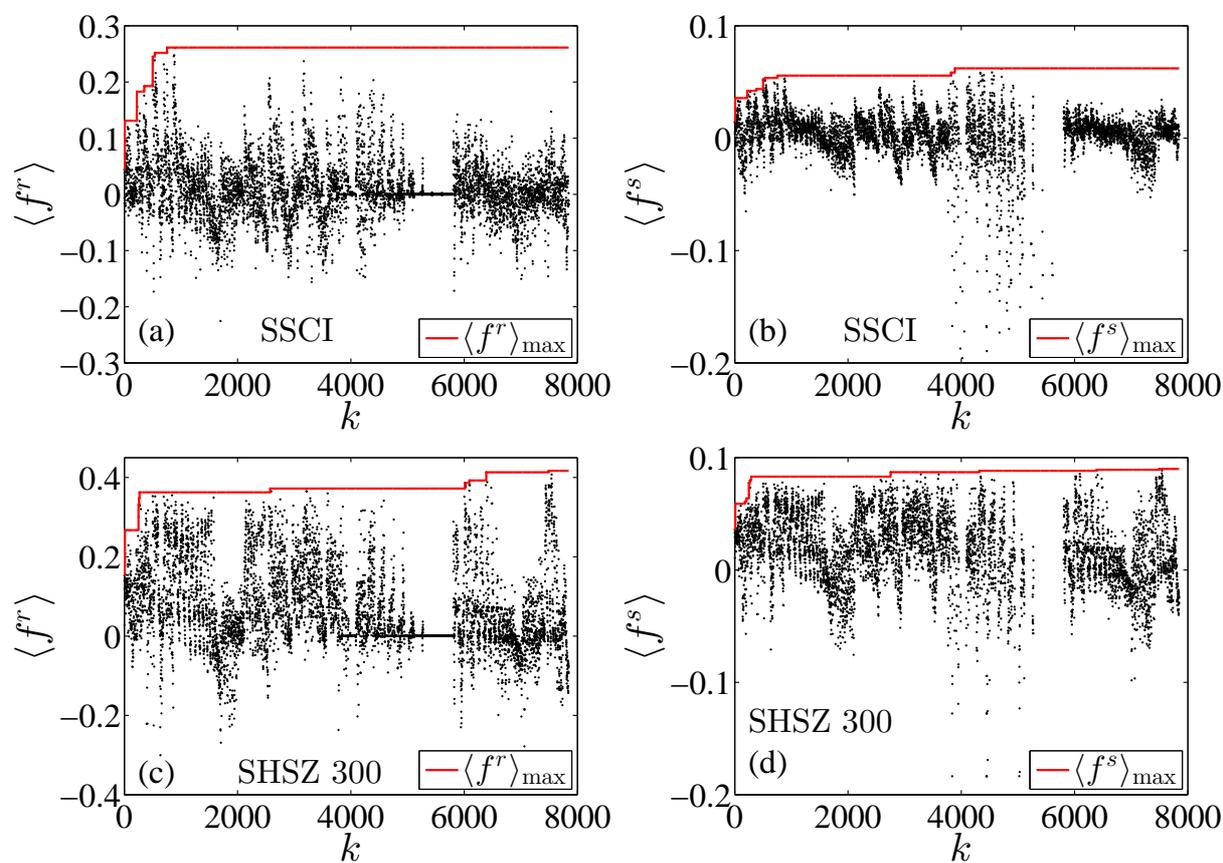}
\caption{Plots of the performance statistics on SSEC (a, b) and SHSZ 300 (c, d) in the whole sample period for all the trading rules in the trading technical strategy pools. (a, c) Performance statistics based on the returns. (b, d) Performance statistics based on the Sharpe ratio.}
\label{Fig:performance} 
\end{figure}

To check whether the best trading rule in the strategy pool can provide significant performance statistics (return or Sharpe ratio) under the consideration of the data snooping effect, we perform back tests on two important indexes (SSCI and SHSZ 300) in Chinese stock markets. We first evaluate the performance of the trading rules on the two indexes in the whole sample period. Figure~\ref{Fig:performance} (a) illustrates the performance statistics based on the returns for each trading rule on SSCI. The dots represent the performance statistic $\langle f^r \rangle$ for each trading rule. We simply set the performance statistic value to be zero for the trading rules that generate no signal. Note that the value of $\langle f^r \rangle$ is annualized by multiplying $252$. The solid line stands for the maximum value of performance statistics achieved by testing the trading rules one by one. One can observe that the final maximum value is $26.13\%$ generated by the $2$ and $20$-day moving averages. Figure~\ref{Fig:performance} (b) illustrates the performance statistics based on the Sharpe ratio for each trading rule on SSCI. The performance statistics $\langle f^s \rangle$ are plotted as dots for each trading rule. Note that the trading rules with zero return performance statistics are excluded. We also do not show 22 Sharpe ratio performance statistics whose values are less than $-0.2$, since we are more interested in the large and position values. The maximum values $\langle f^s \rangle$ can be achieved as $0.0622$, generated by the channel breakouts rule with the number of days of forming a channel, the band and the number of holding days to be 20, 10\% and 5 respectively. Figure ~\ref{Fig:performance} (c, d) illustrates the performance statistics based on the returns and Sharpe ratios of the trading rules on the Shanghai Shenzhen 300 Index. For return statistics, the highest value is $41.69\%$. For Sharpe ratio statistics, the largest value is 0.09, and the smallest value is $-0.47$. Both values are generated by the on-balance volume rule with the $5$ and $20$-day on-balance volume average and $5$ signal-valid days.

\begin{figure}[htb]
\centering
\includegraphics[width=16cm]{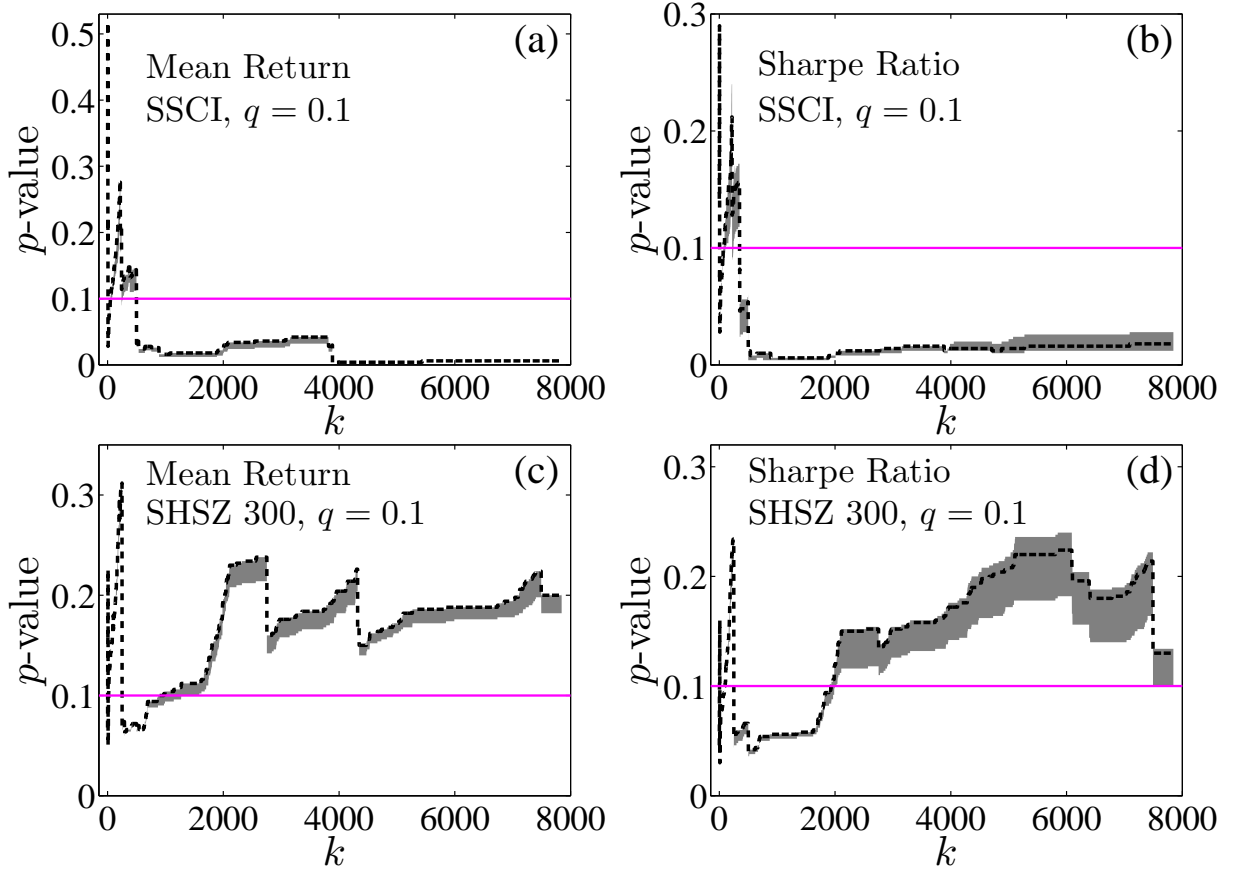}
\caption{Plots of the $p$-value of SPA tests for both performance measurements on both stock indexes. The shadow areas correspond to the upper and lower bounds of the $p$-value. (a) Tests of the performance measurements on the returns on SSCI. (b) Tests of the performance measurements on the Sharpe ratio on SSCI. (c) Tests of the performance measurements on the returns on SHSZ 300. (d) Tests of the performance measurements on the Sharpe ratio on SHSZ 300.}
\label{fig:pvalue} 
\end{figure}

During the procedure of back tests on technical trading strategies, the data-snooping effect that may occur for the testing series is repeatedly used. To correct for this data-snooping effect, we adopt the SPA test to check whether the predictive ability of the best trading rule in the strategy pool is true or just from luck. Figure~\ref{fig:pvalue} illustrates the $p$-value of SPA tests for performance measurements on both indexes with the block shuffling parameter to be $q=0.1$. The shadow areas correspond to the upper and lower bounds of the $p$-value. The dash lines represent the $p$-value of the tested trading rules which were picked up one by one from the strategy pool. Plots of other values of $q$ exhibit very similar patterns and we do not show them here. We can see that at the significance level of 0.1, the tests of both performance measurements on SSCI can be definitely rejected, indicating that technical trading rules have predictive ability for SSCI. This result is further consolidated by the $p$-values listed in the first row of Panel A in Table~\ref{Tb:Return} and \ref{Tb:SharpeRatio}. However, the rejection of the tests on SHSZ 300 is failed for both performance measurements, as shown in Fig.~\ref{fig:pvalue} (b) and the first row of Panel B in Table~\ref{Tb:Return} and \ref{Tb:SharpeRatio}, which means the same trading strategies are ineffective for SHSZ 300.

\setlength\tabcolsep{6pt}
\begin{table*}[htb]
\centering
 \caption{\label{Tb:Return} Performance based on the return of the best trading rules and $p$-value of the SPA tests obtained by applying the trading rules on SSCI and SHSZ 300 in different time periods. The value of mean return is the annualized by multiplying 252. The superscripts $*$, $**$, and $***$ represent the significance level of 10\%, 5\%, and 1\%.}
 \medskip
 \centering
\begin{tabular}{cr@{.}lr@{.}lr@{.}lr@{.}lr@{.}lr@{.}lr@{.}lr@{.}l}
\toprule
Sample &\multicolumn{2}{c}{Return} & \multicolumn{14}{c}{$p$-value}\\
\cline{4-17}
period &\multicolumn{2}{c}{$\langle f^r \rangle_{\max}$}& \multicolumn{2}{l}{$q=0.01$} &\multicolumn{2}{l}{$0.02$}&\multicolumn{2}{l}{$0.05$}& \multicolumn{2}{l}{$0.1$} &\multicolumn{2}{l}{$0.2$}&\multicolumn{2}{l}{$0.5$}& \multicolumn{2}{l}{$1$}\\
\hline
\multicolumn{17}{l}{Panel A: SSCI}\\
19920521-20130630&26&13\%&{\bf{0}}&{\bf{00}}$^{***}$&{\bf{0}}&{\bf{00}}$^{***}$&{\bf{0}}&{\bf{00}}$^{***}$&{\bf{0}}&{\bf{01}}$^{***}$&{\bf{0}}&{\bf{01}}$^{***}$&{\bf{0}}&{\bf{01}}$^{***}$&{\bf{0}}&{\bf{00}}$^{***}$\\
19920521-20050407&26&39\%&{\bf{0}}&{\bf{03}}$^{**}$&{\bf{0}}&{\bf{03}}$^{**}$&{\bf{0}}&{\bf{10}}$^{*}$&0&12&0&16&0&14&0&25\\
20050408-20130630&39&75\%&{\bf{0}}&{\bf{07}}$^{*}$&0&12&0&15&0&20&0&15&{\bf{0}}&{\bf{10}}$^{*}$&{\bf{0}}&{\bf{06}}$^{*}$\\
19920521-19961231&78&27\%&{\bf{0}}&{\bf{01}}$^{***}$&{\bf{0}}&{\bf{10}}$^{*}$&0&32&0&38&0&36&0&32&0&21\\
19930101-19971231&71&23\%&{\bf{0}}&{\bf{03}}$^{**}$&{\bf{0}}&{\bf{08}}$^{*}$&0&23&0&28&0&33&0&28&0&14\\
19940101-19981231&39&81\%&0&14&0&43&0&66&0&76&0&75&0&75&0&75\\
19950101-19991231&37&45\%&{\bf{0}}&{\bf{00}}$^{***}$&{\bf{0}}&{\bf{03}}$^{**}$&0&11&0&18&0&22&0&38&0&47\\
19960101-20001231&30&39\% &{\bf{0}}&{\bf{01}}$^{***}$&{\bf{0}}&{\bf{04}}$^{**}$&0&22&0&43&0&54&0&48&0&41\\
19970101-20011231&28&76\% &{\bf{0}}&{\bf{01}}$^{***}$&{\bf{0}}&{\bf{06}}$^{*}$&0&12&0&28&0&37&0&38&0&25\\
19980101-20021231&30&48\%&{\bf{0}}&{\bf{03}}$^{**}$&0&13&0&26&0&43&0&54&0&50&0&40\\
19990101-20031231&22&93\% &{\bf{0}}&{\bf{05}}$^{**}$&{\bf{0}}&{\bf{06}}$^{**}$&0&11&0&22&0&28&0&29&0&31\\
20000101-20041231&23&14\%&{\bf{0}}&{\bf{03}}$^{**}$&{\bf{0}}&{\bf{10}}$^{*}$&0&39&0&60&0&69&0&73&0&29\\
20010101-20051231&20&57\%&{\bf{0}}&{\bf{06}}$^{*}$&0&12&0&51&0&74&0&78&0&72&0&69\\
20020101-20061231&36&15\%&0&18&0&22&0&44&0&68&0&12&{\bf{0}}&{\bf{07}}$^{*}$&{\bf{0}}&{\bf{03}}$^{**}$\\
20030101-20071231&52&42\%&{\bf{0}}&{\bf{04}}$^{**}$&{\bf{0}}&{\bf{08}}$^{*}$&0&16&0&15&0&12&{\bf{0}}&{\bf{07}}$^{*}$&{\bf{0}}&{\bf{07}}$^{*}$\\
20040101-20081231&68&40\%&{\bf{0}}&{\bf{02}}$^{**}$&{\bf{0}}&{\bf{03}}$^{**}$&{\bf{0}}&{\bf{07}}$^{*}$&0&12&{\bf{0}}&{\bf{09}}$^{*}$&{\bf{0}}&{\bf{08}}$^{*}$&{\bf{0}}& {\bf{05}}$^{**}$\\
20050101-20091231&67&47\%&{\bf{0}}&{\bf{00}}$^{***}$&{\bf{0}}&{\bf{01}}$^{***}$&{\bf{0}}&{\bf{03}}$^{**}$&{\bf{0}}&{\bf{05}}$^{**}$&{\bf{0}}&{\bf{07}}$^{*}$&{\bf{0}}&{\bf{03}}$^{**}$&{\bf{0}}&{\bf{00}}$^{***}$\\
20060101-20101231&53&80\% &{\bf{0}}&{\bf{03}}$^{**}$&{\bf{0}}&{\bf{09}}$^{*}$&0&21&0&30&0&33&0&23&0&13\\
20070101-20111231&44&14\% &{\bf{0}}&{\bf{04}}$^{**}$&0&12&0&12&0&12&0&22&0&36&0&35\\
20080101-20121231&27&81\% &0&11&0&39&0&71&0&73&0&61&0&55&0&38\\
20090101-20130630&29&85\%&{\bf{0}}&{\bf{02}}$^{**}$&{\bf{0}}&{\bf{03}}$^{**}$&0&15&0&32&0&41&0&30&0&25\\

\hline
\multicolumn{17}{l}{Panel B: SZSH 300}\\
20050408-20130630&41&69\%&0&11&0&16&0&19&0&20&0&18&0&14&{\bf{0}}&{\bf{10}}$^{*}$\\
20050408-20091231&79&18\%&{\bf{0}}&{\bf{00}}$^{***}$&{\bf{0}}&{\bf{00}}$^{***}$&{\bf{0}}&{\bf{02}}$^{**}$&{\bf{0}}&{\bf{03}}$^{**}$&{\bf{0}}&{\bf{05}}$^{**}$&{\bf{0}}&{\bf{04}}$^{**}$&{\bf{0}}&{\bf{01}}$^{***}$\\
20060101-20101231&52&57\% &{\bf{0}}&{\bf{05}}$^{**}$&{\bf{0}}&{\bf{08}}$^{*}$&0&22&0&32&0&38&0&30&0&18\\
20070101-20111231&46&22\% &{\bf{0}}&{\bf{08}}$^{*}$&{\bf{0}}&{\bf{10}}$^{*}$&0&16&0&22&0&36&0&28&0&12\\
20080101-20121231&30&97\%&{\bf{0}}&{\bf{06}}$^{*}$&0&16&0&49&0&59&0&67&0&65&0&55\\
20090101-20130630&29&32\%&{\bf{0}}&{\bf{03}}$^{**}$&0&11&0&51&0&65&0&64&0&56&0&40\\
\bottomrule
\end{tabular}
\end{table*}

In order to understand why the same trading strategies give absolutely different results by applying on SSCI and SHSZ 300, which are highly correlated and share very similar behaviors, we first check whether these consequences are attributed to the differences between the data spanning periods. By separating the SSCI indexes at the time point April 8, 2005, which results in two sub-series and the latter series spans the same period as SHSZ 300, the same trading rule testing procedures are carried out on both sub-series. For the performance measurement based on returns, we find that there are 3 $p$-values (corresponding to $q=0.01$, $0.02$ and $0.05$) less than 0.1 for the first sub-series, as listed in the second column of Panel A in Table \ref{Tb:Return} and there is only one $p$-value (corresponding to $q=0.01$) less than 0.1 for the second subseries, as in the third column of Panel A in Table \ref{Tb:Return}. For the performance measurement based the Sharpe ratio, one can observe that only the $p$-value of $q=0.01$ are less than 0.1 for the first subseries, as reported in the second column of Panel A in Table \ref{Tb:SharpeRatio} and the $p$-values of $q=0.01$ and $q=0.02$ are less than 0.1 for the second sub series, as shown in the third column of Panel A in Table \ref{Tb:SharpeRatio}. In some sense, the results of the second sub-series are consistent with those of SHSZ 300. As our benchmark is out of market, we conjecture that the trading rules can make profits only when the price trajectories are in the rising and declining regions. Compared with the whole series of SSCI, both sub-series do not exhibit such prominent up trend behaviors. This is why the testing results two sub-series of SSCI and SHSZ 300 are not significant for all the values of $q$.

\setlength\tabcolsep{6pt}
\begin{table*}[htb]
\centering
 \caption{\label{Tb:SharpeRatio} Performance based on the Sharpe ratio of the best trading rules and $p$-value of the SPA tests obtained by applying the trading rules on SSCI and SHSZ 300 in different time periods. The superscripts $*$, $**$, and $***$ represent the significance level of 10\%, 5\%, and 1\%.}
 \medskip
 \centering
\begin{tabular}{cr@{.}lr@{.}lr@{.}lr@{.}lr@{.}lr@{.}lr@{.}lr@{.}l}
\toprule
Sample &\multicolumn{2}{c}{Sh. Rat.} & \multicolumn{14}{c}{$p$-value}\\
\cline{4-17}
period &\multicolumn{2}{c}{$\langle f^s \rangle_{\max}$}& \multicolumn{2}{l}{$q=0.01$} &\multicolumn{2}{l}{$0.02$}&\multicolumn{2}{l}{$0.05$}& \multicolumn{2}{l}{$0.1$} &\multicolumn{2}{l}{$0.2$}&\multicolumn{2}{l}{$0.5$}& \multicolumn{2}{l}{$1$}\\
\hline
\multicolumn{17}{l}{Panel A: SSCI}\\
19920521-20130630&0&0622&{\bf{0}}&{\bf{00}}$^{***}$&{\bf{0}}&{\bf{00}}$^{***}$&{\bf{0}}&{\bf{01}}$^{***}$&{\bf{0}}&{\bf{02}}$^{**}$&{\bf{0}}&{\bf{03}}$^{**}$&{\bf{0}}&{\bf{2}}$^{**}$&{\bf{0}}&{\bf{02}}$^{**}$\\
19920521-20050407
&0&0570&{\bf{0}}&{\bf{08}}$^{**}$&0&11&0&22&0&31&0&40&0&43&0&55\\
20050408-20130630&0&0917&{\bf{0}}&{\bf{05}}$^{**}$&{\bf{0}}&{\bf{09}}$^{*}$&0&13&0&13&0&13&0&14&{\bf{0}}&{\bf{10}}$^{*}$\\
19920521-19961231&0&1143&{\bf{0}}&{\bf{01}}$^{***}$&{\bf{0}}&{\bf{4}}$^{***}$&0&21&0&33&0&35&0&31&0&22\\
19930101-19971231& 0&1025&{\bf{0}}&{\bf{03}}$^{**}$&0&16&0&32&0&35&0&40&0&37&0&28\\
19940101-19981231& 0&0828&0&29&0&43&0&71&0&72&0&72&0&82&0&82\\
19950101-19991231&0&0947&{\bf{0}}&{\bf{01}}$^{***}$&{\bf{0}}&{\bf{04}}$^{**}$&0&14&0&25&0&33&0&52&0&59\\
19960101-20001231&0&0884 &{\bf{0}}&{\bf{01}}$^{***}$&{\bf{0}}&{\bf{05}}$^{*}$&0&28&0&57&0&72&0&79&0&77\\
19970101-20011231&0&1000  &{\bf{0}}&{\bf{05}}$^{**}$&{\bf{0}}&{\bf{09}}$^{*}$&0&26&0&46&0&58&0&56&0&44\\
19980101-20021231&0&0952&{\bf{0}}&{\bf{04}}$^{**}$&0&22&0&44&0&58&0&69&0&67&0&62\\
19990101-20031231&0&1038 &{\bf{0}}&{\bf{05}}$^{**}$&{\bf{0}}&{\bf{09}}$^{**}$&0&17&0&31&0&37&0&40&0&43\\
20000101-20041231&0&0867 &{\bf{0}}&{\bf{03}}$^{**}$&0&12&0&48&0&74&0&77&0&82&0&79\\
20010101-20051231&0&0775&{\bf{0}}&{\bf{05}}$^{*}$&0&17&0&72&0&93&0&93&0&95&0&95\\
20020101-20061231&0&1303&0&41&0&20&0&81&0&85&0&15&{\bf{0}}&{\bf{10}}$^{*}$&{\bf{0}}&{\bf{05}}$^{**}$\\
20030101-20071231&0&1345&{\bf{0}}&{\bf{02}}$^{**}$&{\bf{0}}&{\bf{06}}$^{*}$&0&13&0&12&{\bf{0}}&{\bf{09}}$^{*}$&{\bf{0}}&{\bf{04}}$^{**}$&{\bf{0}}&{\bf{04}}$^{**}$\\
20040101-20081231&0&1397&{\bf{0}}&{\bf{01}}$^{***}$&{\bf{0}}&{\bf{05}}$^{**}$&{\bf{0}}&{\bf{05}}$^{**}$&{\bf{0}}&{\bf{09}}$^{*}$&{\bf{0}}&{\bf{06}}$^{*}$&{\bf{0}}&{\bf{06}}$^{*}$&{\bf{0}}& {\bf{03}}$^{**}$\\
20050101-20091231&0&1395&{\bf{0}}&{\bf{00}}$^{***}$&{\bf{0}}&{\bf{01}}$^{***}$&{\bf{0}}&{\bf{02}}$^{**}$&{\bf{0}}&{\bf{03}}$^{**}$&{\bf{0}}&{\bf{05}}$^{**}$&{\bf{0}}&{\bf{03}}$^{**}$&{\bf{0}}&{\bf{02}}$^{**}$\\
20060101-20101231&0&1101 &{\bf{0}}&{\bf{04}}$^{**}$&{\bf{0}}&{\bf{07}}$^{*}$&0&13&0&20&0&25&0&32&0&26\\
20070101-20111231&0&1094 &0&17&0&11&0&11&0&12&0&20&0&32&0&30\\
20080101-20121231&0&0882 &0&16&0&52&0&69&0&81&0&83&0&82&0&79\\
20090101-20130630&0&1089 &{\bf{0}}&{\bf{00}}$^{***}$&{\bf{0}}&{\bf{07}}${*}$&0&36&0&52&0&58&0&53&0&46\\
\hline
\multicolumn{17}{l}{Panel B: SZSH 300}\\
20050408-20130630& 0&0900&0&11&0&14&0&13&0&13&0&15&0&16&0&14\\
20050408-20091231&0&1503&{\bf{0}}&{\bf{00}}$^{***}$&{\bf{0}}&{\bf{00}}$^{***}$&{\bf{0}}&{\bf{02}}$^{**}$&{\bf{0}}&{\bf{04}}$^{**}$&{\bf{0}}&{\bf{06}}$^{*}$&{\bf{0}}&{\bf{04}}$^{**}$&{\bf{0}}&{\bf{01}}$^{***}$\\
20060101-20101231&0&1086 &{\bf{0}}&{\bf{02}}$^{**}$&{\bf{0}}&{\bf{07}}$^{*}$&0&23&0&26&0&32&0&35&0&31\\
20070101-20111231&0&1076 &{\bf{0}}&{\bf{06}}$^{*}$&{\bf{0}}&{\bf{10}}$^{*}$&0&15&0&18&0&29&0&33&0&26\\
20080101-20121231&0&0829&0&12&0&37&0&68&0&80&0&88&0&92&0&90\\
20090101-20130630&0&0897&{\bf{0}}&{\bf{05}}$^{**}$&0&18&0&54&0&78&0&81&0&83&0&89\\
\bottomrule
\end{tabular}
\end{table*}

To further check the conjecture that our trading rules can only earn money in the market with prominent trends, we perform the trading rule tests on both indexes with a five-year moving window. As shown in Tables~\ref{Tb:Return} and \ref{Tb:SharpeRatio}, all of the series in the windows of 2005-2009 pass the SPA tests based on the return and the Sharpe ratio performance measurements at the 0.1 significance level for different values of $q$. During that period, SSCI rose from 998 to 6124 and a bubble was diagnosed from mid-2005 to October 2007 \cite{Jiang-Zhou-Sornette-Woodard-Bastiaensen-Cauwels-2010-JEBO}. For the other periods, no such obviously rising trends are observed, which gives rise to the fact that not all $p$-values for the different values of $q$ are less than 0.1. From our testing results, we also find that the values of the block shuffling parameter $q$ does have influences on the significance of the SPA tests. By increasing $q$ from 0.01 to 1, the number of tests (including all time windows for both indexes) whose hypothesises are rejected exhibit a first decreasing then increasing pattern, where the lowest number is located at around $q=0.1$.

\section{Conclusion}

In this paper, we use the SPA test \cite{Hansen-2005-JBES} to demonstrate how to correct for the influence of the data-snooping effect when one tries to look for profitable trading rules from a pool of trading strategies by repeatedly using a given financial price series. Taking SSCI and SHSZ 300 as examples, we find that technical trading rules can make reliable profits in SSCI while losing the profitable ability in SHSZ 300. This contradicting consequence can be attributed to the fact that there is a more clear price trend in SSCI, which is the base of technical trading rules. By applying the trading rules on the sub-series of SSCI with the same spanning period as that of SHSZ 300, we find that the efficiency of the technical trading rules is greatly reduced. From the back tests in each moving window with a size of five years, we find that periods which contain extreme upward or downward trends such as financial bubbles, offer good opportunities for investors to make profits by the adoption of technical trading strategies.

\section*{Acknowledgements}

This work was partially supported by the National Natural Science Foundation of China (11075054 and 71131007), Shanghai ``Chen Guang'' Project (2012CG34), Program for Changjiang Scholars and Innovative Research Team in University (IRT1028), and the Fundamental Research Funds for the Central Universities.


\begin{thebibliography}{10}
\expandafter\ifx\csname url\endcsname\relax
  \def\url#1{\texttt{#1}}\fi
\expandafter\ifx\csname urlprefix\endcsname\relax\def\urlprefix{URL }\fi
\expandafter\ifx\csname href\endcsname\relax
  \def\href#1#2{#2} \def\path#1{#1}\fi

\bibitem{Treynor-Ferguson-1985-JF}
J.-L. Treynor, R.~Ferguson, {In defense of technical analysis}, J. Finance
  40~(3) (1985) 757--773.

\bibitem{Brock-Lalonishok-Lebaron-1992-JF}
W.~Brock, J.~Lakonishok, B.~Lebaron, {Simple technical trading rules and the
  stochastic properties of stock returns}, J. Finance 47 (1992) 1731--1764.

\bibitem{Neely-Weller-1999-JIMF}
C.~J. Neely, P.~A. Weller, {Technical trading rules in the european monetary
  system}, J. Int. Money Finance 18 (1999) 429--458.

\bibitem{Lucke-2003-AE}
B.~Lucke, {Are technical trading rules profitable? Evidence for
  head-and-shoulder rules}, Appl. Econ. 35 (2003) 33--40.

\bibitem{Anderson-Faff-2005-ARJ}
J.~Anderson, R.~Faff, {Profitability of trading rules in futures markets},
  Accounting Research Journal 18 (2005) 83--92.

\bibitem{Kung-2009-RAE}
J.~J. Kung, {Predictability of technical trading rules: Evidence from the
  Taiwan Stock Market}, Review of Applied Economics 5 (2009) 49--65.

\bibitem{Fang-Jacobsen-Qin-2014-RFE}
J.~L. Fang, B.~Jacobsen, Y.~F. Qin, {Predictability of the simple technical
  trading rules: An out-of-sample test}, Rev. Financial Econ. 23 (2014) 30--45.

\bibitem{Yu-Nartea-Gan-Yao-2013-IREF}
H.~Yu, G.-V. Nartea, C.~Gan, L.-J. Yao, {Predictive ability and profitability
  of simple technical trading rules: Recent evidence from Southeast Asian stock
  markets}, Int. Rev. Econ. Finance 25 (2013) 356--371.

\bibitem{Hudson-Dempsey-Keasey-1996-JBF}
R.~Hudson, M.~Dempsey, K.~Keasey, {A note on the weak form efficiency of
  capital markets: The application of simple technical trading rules to UK
  stock prices 1935-1994}, J. Bank. Finance 20 (1996) 1121--1132.

\bibitem{White-2000-Em}
H.~White, {A reality check for data snooping}, Econometrica 68~(5) (2000)
  1097--1126.

\bibitem{Jensen-Bennington-1970-JF}
M.-C. Jensen, G.-A. Bennington, {Random walks and technical theories: Some
  additional evidence}, J. Finance 25~(2) (1970) 469--482.

\bibitem{Lo-MacKinlay-1990-RFS}
A.-W. Lo, A.-C. MacKinlay, {Data-snooping biases in tests of financial asset
  pricing models}, Rev. Financ. Stud. 3~(3) (1990) 431--467.

\bibitem{Sullivan-Timmermann-White-1999-JF}
R.~Sullivan, A.-G. Timmermann, H.~White, {Data-Snooping, Technical Trading Rule
  Performance and the Bootstrap}, J. Finance LIV~(5) (1999) 1647--1691.

\bibitem{Hansen-2005-JBES}
P.-R. Hansen, {A test for superior predictive ability}, J. Bus. Econ. Stat.
  23~(4) (2005) 365--380.

\bibitem{Park-Irwin-2010-JFutM}
C.-H. Park, S.-H. Irwin, {A reality check on technical trading rule profits in
  U.S. futures markets}, J. Fut. Markets 30 (2010) 633--659.

\bibitem{Shynkevich-2012-JBF}
A.~Shynkevich, {Performance of technical analysis in growth and small cap
  segments of the US equity market}, J. Bank. Finance 36 (2012) 193--208.

\bibitem{Chen-Huang-Lai-2009-JAsianE}
C.-W. Chen, C.-S. Huang, H.-W. Lai, {The impact of data snooping on the testing
  of technical analysis: An empirical study of Asian stock markets}, J. Asian
  Econ. 20 (2009) 580--591.

\bibitem{Fama-Blume-1965-JB}
E.-F. Fama, M.-E. Blume, {Filter rules and stock-market trading}, J. Business
  39~(1) (1966) 226--241.

\bibitem{Politis-Romano-1994-JASA}
D.-N. Politis, J.-P. Romano, {The stationary bootstrap}, J. Am. Stat. Assoc. 89
  (1994) 1303--1313.

\bibitem{Jiang-Zhou-Sornette-Woodard-Bastiaensen-Cauwels-2010-JEBO}
Z.-Q. Jiang, W.-X. Zhou, D.~Sornette, R.~Woodard, K.~Bastiaensen, P.~Cauwels,
  {Bubble diagnosis and prediction of the 2005-2007 and 2008-2009 Chinese stock
  market bubbles}, J. Econ. Behav. Org. 74 (2010) 149--162.
\newblock \href {http://dx.doi.org/10.1016/j.jebo.2010.02.007}
  {\path{doi:10.1016/j.jebo.2010.02.007}}.

\end{thebibliography}

\end{document}